\documentstyle[aps]{revtex}

\begin{document}

\author{Thomas W. Kephart \footnote[1]{E-mail: kephart@vanderbilt.edu} }
\title{Classification of $SUSY$ and non--$SUSY$ Chiral Models from Abelian Orbifolds 
$AdS/CFT$}
\date{\today}
\vskip 10pt
\address{Department of Physics and Astronomy,\\
Vanderbilt University, Nashville, TN 37325;\\
}
\author{ Heinrich P\"{a}s \footnote[2]{E-mail: paes@physik.uni-wuerzburg.de}}
\address{
Institut f\"ur Theoretische Physik und Astrophysik\\
Universit\"at W\"urzburg\\ D-97074 W\"urzburg, Germany}

\maketitle

\bigskip  

\begin{abstract}

We classify compactifications of the type $IIB$
superstring on
$AdS_{5}\times S^{5}/\Gamma $, where $\Gamma $ is an abelian  
group of order $n\leq 12$.  
Appropriate embedding of $\Gamma$ in the isometry of $S^5$
yields both $SUSY$ and non--$SUSY$ chiral models that can contain the
minimal $SUSY$ standard model or the standard model.
New non-SUSY three family models with $\Gamma=Z_8$ are introduced, 
which lead to the right Weinberg angle for TeV trinification.
 
\end{abstract}


\bigskip
  
\newpage

\section{Introduction}

When one bases models on conformal field theory gotten from the large $N$
expansion of the $AdS/CFT$ correspondence \cite{Maldacena:1998re}, stringy
effects can arise at an energy scale as low as a few TeV. These 
models can potentially test string theory and examples with low energy scales are known
in orbifolded $AdS_{5}\times S^{5}$. The first three-family model of this
type had ${\cal N}=1$ $SUSY$ and was based on a $Z_{3}$ orbifold \cite
{Kachru:1998ys}, see also \cite{Kephart:2001qu}. However, since then some of the
most studied examples have been models without supersymmetry based on both
abelian \cite{Frampton:1999wz}, \cite{Frampton:1999nb}, \cite
{Frampton:1999ti} and non-abelian \cite{Frampton:2000zy}, \cite
{Frampton:2000mq} orbifolds of $AdS_{5}\times S^{5}$. Recently both SUSY and nonSUSY three family $Z_{12}$ 
orbifold models \cite{Frampton:2002st,Frampton:2003jx}
have been shown to unify at a low scale ($\sim$ 4 TeV) and to have promise of testability. One motivation for
studying the non--$SUSY$ case is that the need for supersymmetry is less
clear as: (1) the hierarchy problem is absent or ameliorated
\footnote{Compare however the discussion in \cite{hier}.}, (2)
the difficulties involved in breaking
the remaining ${\cal N}=1$ $SUSY$ can be avoided if the orbifolding already
results in ${\cal N}=0$ $SUSY$, and (3) many of the
effects of $SUSY$ are still present in the theory, just hidden. For example,
the bose-fermi state count matches, RG equations preserve vanishing $\beta $
functions to some number of loops, etc. Here we concentrate on abelian
orbifolds with and without supersymmetry, where the orbifolding group $\Gamma$ has
order $n=o(\Gamma)\leq 12$. We systematically study those
cases with chiral matter ($i.e.$, in the $SUSY$ case, those with an imbalance between chiral
supermultiplets and anti-chiral supermultiplets, and in the non--$SUSY$ case with a net
imbalance between left and right handed fermions). We find all chiral models for 
$n\leq 12$. Several of these contain the standard model ($SM$)
or the minimal supersymmetric standard model ($MSSM$) with three or four families.

We begin with a summary of how orbifolded $AdS_{5}\times S^{5}$ models are
constructed (for more details see \cite{Frampton:2000mq}). First we select a
discrete subgroup $\Gamma $ of the $SO(6)\sim SU(4)$ isometry of $S^{5}$
with which to form the orbifold $AdS_{5}\times S^{5}/\Gamma $. The
replacement of $S^{5}$ by $S^{5}/\Gamma $ reduces the supersymmetry to $%
{\cal N}=$ 0, 1 or 2 from the initial ${\cal N}=4$, depending on how $\Gamma 
$ is embedded in the isometry of $S^{5}$. The cases of interest here are $%
{\cal N}=0$ and ${\cal N}=1\ $ $SUSY$ where $\Gamma $ embeds irreducibly in the $%
SU(4)$ isometry or in an $SU(3)$ subgroup of the $SU(4)$ isometry, respectively. $I.e.$, 
to achieve ${\cal N}=0$ we embed rep($\Gamma )\rightarrow {\bf 4}$ of $%
SU(4)$ as ${\bf 4}=({\bf r})$ where ${\bf r}$ is a nontrivial  
four dimensional representation of $\Gamma ;$ for ${\cal N}=1$ we
embed rep($\Gamma )\rightarrow {\bf 4}$ of $SU(4)$ as ${\bf 4}=({\bf 1},{\bf %
r})$ where ${\bf 1}\ $is the trivial 
irreducible representation
(irrep) of $\Gamma $ and ${\bf r}$ is a
nontrivial three dimensional representation of $%
\Gamma .$

For ${\cal N}=0$ the fermions are given by 
$\sum_{i}{\bf 4}\otimes R_{i}$
and the scalars by 
$\sum_{i}{\bf 6}\otimes R_{i}$
where the set \{$R_{i}\}$ runs over all the irreps of $\Gamma .$ For
$\Gamma$ abelian, the irreps are all one dimensional and as a
consequence of the choice of $N$ in the $1/N$ expansion, the gauge group 
\cite{Lawrence:1998ja} is $SU^{n}(N)$. In the ${\cal N}=1\ $ $SUSY$ case,
chiral supermultiples generated by this embedding are given by 
$ \sum_{i}
{\bf 4}\otimes R_{i}$
where again \{$R_{i}\}$ runs over all the 
(irreps) of $\Gamma .$ Again for abelian $\Gamma$, the irreps
are all one dimensional and the gauge group is again $SU^{n}(N)$. Chiral
models require the {\bf 4} to be complex (${\bf 4}\neq {\bf 4}^{*})$ while a
proper embedding requires ${\bf 6}={\bf 6}^{*}$ where ${\bf 6}$=(${\bf 4}%
\otimes {\bf 4})_{antisym}$. (Even though the ${\bf 6}$ does not enter the
model in the ${\cal N}=1\ $ $SUSY$ case, mathematical consistency requires $
{\bf 6}={\bf 6}^{*}$, see \cite{Frampton:2003vc}.)

We now have the required background to begin building chiral models. We
choose $N=3$ throughout. If $SU_{L}(2)$ and $U_{Y}(1)$ are embedded in
diagonal subgroups $SU^{p}(3)$ and $SU^{q}(3)$ respectively, of the initial $
SU^{n}(3)$, the ratio $\frac{\alpha _{2}}{\alpha _{Y}}$ is $\frac{p}{q},$
leading to a calculable initial value of $\theta _{W}$ with, 
$ \sin^{2}\theta _{W}=\frac{3}{3+5\left( \frac{p}{q}\right) }.$
The more standard approach is to break the initial $SU^{n}(3)$ to $%
SU_{C}(3)\otimes SU_{L}(3)\otimes SU_{R}(3)$ where $SU_{L}(3)$ and $%
SU_{R}(3) $ are embedded in diagonal subgroups $SU^{p}(3)$ and $SU^{q}(3)$
of the initial $SU^{n}(3)$. We then embed all of $SU_{L}(2)$ in $SU_{L}(3)$
but $\frac{1}{3}$ of $U_{Y}(1)$ in $SU_{L}(3)$ and the other $\frac{2}{3}$
in $SU_{R}(3)$. This modifies the $\sin ^{2}\theta _{W}$ formula to: 
$\sin ^{2}\theta _{W}=\frac{3}{3+5\left( \frac{\alpha _{2}}{\alpha _{Y}}
\right) } =\frac{3}{3+5\left( \frac{3p}{p+2q}\right) }$, which coincides 
with the previous result when $p=q$. One should use the second (standard) embedding 
when calculating $\sin ^{2}\theta _{W}$ for any of the models obtained below. A similar relation
holds for Pati--Salam type models \cite{Frampton:2001xh} and their generalizations \cite{Kephart:2001ix}, but this would
require investigation of models with $N\geq 4$ which are not included in this study.
Note, if $\Gamma=Z_n$ the initial ${\cal N}=0$ orbifold
model (before any symmetry breaking) is completely fixed (recall we always
are taking $N=3$) by the choice of $n$ and the embedding ${\bf 4=}%
(\alpha ^{i},\alpha ^{j},\alpha ^{k},\alpha ^{l}),$ so we define these models
by $M_{ijkl}^{n}.$ The conjugate models $M_{n-i,n-j,n-k,n-l}^{n}$ contain
the same information, so we need not study them separately.

As we have previously studied chiral $\Gamma=Z_n$ models with ${\cal N}=1$ $SUSY,$ we first summarize
those results before 
concentrating on ${\cal N}=0$. At the end we consider both ${\cal N}=1$ and ${\cal N}=0$ 
models where $\Gamma$ is abelian but not a single $Z_n$. For instance $\Gamma=Z_3 \times Z_3\neq Z_9$.

\section{Summary of ${\cal N}=1$ chiral $Z_n$ models}

To tabulate the
possible models for each value of $n$, we first show that a
proper embedding ($i.e$., ${\bf 6=6}^{*}$) for ${\bf 4=}({\bf 1},\alpha
^{i},\alpha ^{j},\alpha ^{k})$ results when $i+j+k=n$. To do this we use the fact
that the conjugate model has $i\rightarrow i^{\prime }=n-i,$ $j\rightarrow
j^{\prime }=n-j$ and $k\rightarrow k^{\prime }=n-k.$ Summing we find $%
i^{\prime }+j^{\prime }+k^{\prime }=3n-(i+j+k)=2n.$ But from ${\bf 6}$=(${\bf 4}%
\otimes {\bf 4})_{antisym}$ we find ${\bf 6=}(\alpha ^{i},\alpha ^{j},\alpha
^{k},\alpha ^{j+k},\alpha ^{i+k},\alpha ^{i+j}),$ but $i+j=n-k=k^{\prime }.$
Likewise $i+k=j^{\prime }$ and $j+k=i^{\prime }$ so ${\bf 6=}(\alpha
^{i},\alpha ^{j},\alpha ^{k},\alpha ^{i^{\prime }},\alpha ^{j^{\prime
}},\alpha ^{k^{\prime }})$ and this is ${\bf 6}^{*}$ up to an automorphism
which is sufficient to provide a proper embedding (or to provide real scalars in the
non-SUSY models). Models with $i+j+k=n$
(we will call these partition models) are always chiral, with total
chirality (number of chiral states) $\chi =3N^{2}n$ except in the case where $n$ is even and one of $i$%
, $j$, or $k$ is $n/2$ where $\chi =2N^{2}n.$ (No more than one of $i$, $j$,
and $k$ can be $n/2$ since they sum to $n$ and are all positive.) This
immediately gives us a lower bound on the number of chiral models at fixed $%
n $. It is the number of partitions of $n$ into three non-negative
integers. There is another class of models with $i^{\prime }=k$ and $%
j^{\prime }=2j,$ and total chirality $\chi =N^{2}n;$ for example a $Z_{9}$
orbifold with ${\bf 4=}({\bf 1},\alpha ^{3},\alpha ^{3},\alpha ^{6}).$ And
there are a few other sporadically occurring cases like $M_{124}^{6}$, which
typically have reduced total chirality, $\chi <3N^{2}n$. 
Such "nonpartition'' - i.e. neither partition nor double partition - models
can fail other more subtle constraints on consistent embedding  \cite{Frampton:2003vc},
but we list them here because they have vanishing anomaly coefficients and
vanishing one loop $\beta$ functions, and so are still of phenomenological interest
from the gauge theory model building perspective.


We now list all the ${\cal N}=1$, $Z_{n}$ orbifold models up to $n=12$ along with the
total chirality of each model, (see Table 1).

A systematic
search  through $n \leq 7$ yields four models that can result a in three-family
MSSM. They are $M_{111}^{3}$, $%
M_{122}^{5}, $ $M_{123}^{6},$ and $M_{133}^{7}$. There may be many
more models with sensible phenomenology at larger $n$, and we have given
one example $M_{333}^{9}$, with particularly simple 
spontaneous symmetry breaking, that is also a member of an infinite series of
models $M_{\frac{n}{3}\frac{n}{3}\frac{n}{3}}^{n}$, which all can lead to
three-family $MSSM$s. The value of $\sin ^{2}\theta _{W}$ at $SU^n(3)$ unification
was calculated for all these three family models in \cite{Kephart:2001qu}. 
This completes the summary of 
${\cal N}=1$ chiral $Z_n$ models, so we now proceed to investigate chiral $Z_n$ models
with no remaining supersymmetry.

\section{  ${\cal N}=0 $ chiral $Z_n$ models}

We begin this section by studying the first few ${\cal N}=0 $ chiral $Z_n$ models.
Insights gained here will allow us to generalize and give results to arbitrary $n$.
First, the allowed $\Gamma =Z_{2}$ and $Z_{3},$ ${\cal N}=0$ orbifolds have
only real representations and therefore will not yield chiral models. 
Next, for $\Gamma =Z_{4}$ the choice ${\bf 4}=(\alpha ,\alpha ,\alpha
,\alpha )$ with $N=3$ where $\alpha =e^{\frac{\pi i}{2}}$ (in what follows
we will write $\alpha =e^{\frac{2\pi i}{n}}$ for the roots of unity that
generate $Z_{n})$, yields an $SU^{4}(3)$ chiral model with the fermion content shown
in Table 2.

The scalar content of this model is given in Table 3 and a VEV for say a
(3,1,\={3},1) breaks the symmetry to $SU_{D}(3)\times SU_{2}(3)\times
SU_{4}(3)$ but renders the model vectorlike, and hence uninteresting, so we consider it no further.
The only other choice of embedding is a nonpartition model 
with $\Gamma = Z_4$ is ${\bf 4}=(\alpha ,\alpha ,\alpha ,\alpha ^{3})$ 
but it leads to the same
scalars with half the chiral fermions so we move on to $Z_{5}.$

There is one chiral model for $\Gamma =Z_{5}$ and it is fixed by choosing $%
{\bf 4}=(\alpha ,\alpha ,\alpha ,\alpha ^{2})$, leading to ${\bf 6}=(\alpha
^{2},\alpha ^{2},\alpha ^{2},\alpha ^{3},\alpha ^{3},\alpha ^{3})$ with real
scalars. It is straightforward to write down the particle content of this $%
M_{1112}^{5}$ model. The best one can do toward the construction of the
standard model is to give a VEV to a (3,1,\={3},1,1) to break the $SU^{5}(3)$
symmetry to $SU_{D}(3)\times SU_{2}(3)\times SU_{4}(3)\times SU_{5}(3).$ Now
a VEV for (1,3,\={3},1) completes the breaking to $SU^{3}(3),$ but the only
remaining chiral fermions are $2[(3,\bar{3},1)+(1,3,\bar{3})+(\bar{3},1,3)]$
which contains only two families.

Moving on to $\Gamma =Z_{6}$ we find two models
where, as with the previous $Z_{5}$ model, the ${\bf 4}$ is arranged so that 
${\bf 4=\ }(\alpha ^{i},\alpha ^{j},\alpha
^{k},\alpha ^{l})$ with $i+j+k+l=n.$
These have ${\bf 4}%
=(\alpha ,\alpha ,\alpha ,\alpha ^{3})$ and ${\bf 4}=(\alpha ,\alpha ,\alpha
^{2},\alpha ^{2})$ and were defined as partition
models in \cite{Kephart:2001qu} when $i$ was equal to zero.
Here we generalize and call all models satisfying $i+j+k+l=n$ partition
models. We have now introduced most of the
background and notation we need, so at this point (before completing the investigation of
the $\Gamma =Z_{6}$ models) it is useful to give a
summary (see Table 4) of all ${\cal N}=0$ chiral $Z_{n}$ models with
real ${\bf 6}$'s for $n\leq 12.$ 
We note that the $%
n=8$ partition model with ${\bf 4}=(\alpha ,\alpha ,\alpha ^{2},\alpha ^{4})$
has $\chi /N^{2}=16;$ the other four have $\chi /N^{2}=32.$ Of the nine $%
Z_{10}$ partition models$,$ 2 have $\chi /N^{2}=30$ and the other 7 have $%
\chi /N^{2}=40.$ The $Z_{12}$ partition
models derived from ${\bf 4}=(\alpha ,\alpha ,\alpha ^{4},\alpha ^{6}),$ $%
{\bf 4}=(\alpha ,\alpha ^{2},\alpha ^{3},\alpha ^{6}),$ and ${\bf 4}=(\alpha
^{2},\alpha ^{2},\alpha ^{2},\alpha ^{6})$ have $\chi /N^{2}=36;$ the others
have $\chi /N^{2}=48.$

A new class of models appears in Table 4; these are 
the double partition models. 
They have $i+j+k+l=2n$ and none are
equivalent to single partition models (if we require that $i$, $j$, $k$, and $%
l $ are all positive integers) with $i+j+k+l=n.$ The ${\cal N}=1$
nonpartition models have been classified \cite{Frampton:2003vc}, and we
find eleven ${\cal N}=0$ examples in Table 4. While they have a self
conjugate ${\bf 6,}$ this is only a necessary condition that may be
insufficient to insure the construction of viable string theory based
models \cite{Frampton:2003vc}. 
However, as is the ${\cal N}=1$ case, the ${\cal N}=0$ nonpartition models may still be interesting
phenomenologically and as a testing ground for models with the potential of
broken conformal invariance.  

For $Z_{n}$ orbifold models with $n$ a prime number, only partition models
arise. The non--partition and double partition models only occur when $n$ is not
a prime number, and only a few are independent. Consider $n=12,$ here
we can write $Z_{12}=Z_{4}\times Z_{3}.$ If we write an element of this
group as $\gamma \equiv (a,b),$ where $a$ is a generator of $Z_{4}\ $and $b$
of $Z_{3},$ then $\gamma ^{2}\equiv (a^{2},b^{2}),$ $\gamma ^{3}\equiv
(a^{3},1),$ etc. The full group is generated by any one of the elements $\gamma =(a,b),$ $%
\gamma ^{5}=(a,b^{2}),$ $\gamma ^{7}=(a^{3},b),$ or $\gamma
^{11}=(a^{3},b^{2}).$ The other choices do not faithfully represent the
group. Letting $\alpha =\gamma ^{11}$ give a conjugate model, $e.g$., it
transforms $(\alpha ,\alpha ^{6},\alpha ^{8},\alpha ^{9})$ into $(\gamma
^{11},\gamma ^{6},\gamma ^{4},\gamma ^{3}),$ so this pair of double partition models are equivalent,
while letting $\alpha =\gamma ^{5}$ transforms $(\alpha ,\alpha ^{6},\alpha
^{8},\alpha ^{9})$ into the equivalent model $(\gamma ^{5},\gamma
^{6},\gamma ^{4},\gamma ^{9}),$ and $\alpha =\gamma ^{7}$ transforms $%
(\alpha ,\alpha ^{6},\alpha ^{8},\alpha ^{9})$ into the equivalent model $%
(\gamma ^{7},\gamma ^{6},\gamma ^{8},\gamma ^{3}).$ Hence a systematic use
of these operations on the non--partition and double partition models can
reduce them to the equivalence classes listed in the tables.

It is easy to prove we always have a proper embedding ($i.e$., ${\bf 6=6}%
^{*} $) for the ${\bf 4=\ }(\alpha ^{i},\alpha ^{j},\alpha ^{k},\alpha ^{l})%
{\bf \ }$ when $i+j+k+l=n$ (or $2n$). To show this note from ${\bf 6}$=(${\bf 4}%
\otimes {\bf 4})_{antisym}$ we find ${\bf 6=}\ (\alpha ^{i+j},\alpha
^{i+k},\alpha ^{i+l},\alpha ^{j+k},\alpha ^{j+l},\alpha ^{k+l}),$ but $%
i+j=n-k-l=-(k+l)%
\mathop{\rm mod}%
n,$ $i+k=n-j-l=-(j+l)%
\mathop{\rm mod}%
n,$ and $i+l=n-j-k=-(j+k)%
\mathop{\rm mod}%
n,$ so this gives ${\bf 6=}$ $(\alpha ^{-(k+l)},\alpha ^{-(j+l)},\alpha
^{-(j+k)},\alpha ^{j+k},\alpha ^{j+l},\alpha ^{k+l})={\bf 6}^{*}.$ A simple modification of this proof
also applies to the double partition models.

Now let us return to $\Gamma =Z_{6}$ where the 
partition models of interest are :(1) ${\bf 4}=(\alpha ,\alpha
,\alpha ^{2},\alpha ^{2})$ where one easily sees that VEVs for $(3,1,\bar{3}%
,1,1,1)$ and then $(1,3,\bar{3},1,1)$ lead to at most two families, while
other SSB routes lead to equal or less chirality. (2) ${\bf 4}=(\alpha
,\alpha ,\alpha ,\alpha ^{3})$ where VEVs for $(3,1,\bar{3},1,1,1)$ followed
by a VEV for $(1,3,\bar{3},1,1)$ leads to an $SU^{4}(3)$ model containing
fermions 2[$(3,\bar{3},1,1)+(1,3,\bar{3},1)+(1,1,3,\bar{3})+(\bar{3},1,1,3)]$%
. However, there are insufficient scalars to complete the symmetry breaking
to the standard model. In fact, one cannot even achieve the trinification
spectrum.

The double partition $Z_{6}$ model ${\bf 4}=(\alpha ,\alpha ^{3},\alpha
^{4},\alpha ^{4})$ is relatively complicated, since there are 24 different
scalar representations in the spectrum, and this makes the SSB analysis
rather difficult. We have investigated a number of possible SSB pathways,
but have found none that lead to the SM with at least three families.
However, since our search was not exhaustive, we cannot make a definitive
statement about this model. As stated elsewhere, the non-partition models
are difficult to interpret, if not pathological, so we have not studied the
SSB pathways for these $Z_{6}$ models.

We move on to $Z_{7}$, where there are three partition models: (1) for ${\bf %
4}=(\alpha ,\alpha ^{2},\alpha ^{2},\alpha ^{2})$, we find no SSB pathway to
the SM. There are paths to an SM with less than three families, e. g., VEVs
for $(3,1,1,\bar{3},1,1,1),$ $(1,3,1,\bar{3},1,1),$ $(3,\bar{3},1,1,1),$ and
$(1,3,\bar{3},1)$ lead to one family at the $SU^{3}(3)$ level; 
(2) for $%
{\bf 4}=(\alpha ,\alpha ,\alpha ,\alpha ^{4})$, again we find only paths to
family-deficient standard models. 
An example is where we have VEVs for 
$(3,1,\bar{3},1,1,1,1),$ $(1,3,\bar{3},1,1,1),$ $(3,1,\bar{3},1,1),$ and $%
(1,3,\bar{3},1)$,
which lead to a two-family $SU^{3}(3)$ model; 
(3) finally, ${\bf %
4}=(\alpha ,\alpha ,\alpha ^{2},\alpha ^{3})$ is the model discovered in 
\cite{Frampton:1999wz}, where VEVs to 
$(1, 3,1,\bar{3},1,1,1)$, $(1,1,3,\bar{3},1,1)$, $(1,1,3,\bar{3},1)$
and $(1,1,3,\bar{3})$ 
lead to a three family model with the correct Weinberg angle
at the $Z$-pole, $\sin^2 \theta_W=3/13$.

For $Z_{n}$ with $n\geq 8,$ the number of representations of matter
multiplets has already grown to a degree where it makes a systematic
analysis of the models prohibitively time-consuming. It is thus helpful to
have further motivation to study particular examples or limited sets of
these models with large $n$ values. Thus we searched for examples which 
break
$SU(3)^8$ down 
to diagonal subgroups
$SU(3)^4\times SU(3)^3 \times SU(3)$, 
since this implies the right Weinberg angle for TeV trinification \cite{Glashow:1984gc},
$\sin^2 \theta_W= 
3/13$, when embedding $SU(3)_L$ and $SU(3)_R$ into the diagonal subgroups of
$SU(3)^4$ and $SU(3)$, 
respectively.
There are actually 11 different possibilities
to
break $SU(3)^8$ down 
to $SU(3)^4\times SU(3)^3 \times SU(3)$, assuming the necessary scalars 
exist. While none of these paths was successful for 
${\bf %
4}=(\alpha ,\alpha,\alpha,\alpha ^{5})$, the model ${\bf %
4}=(\alpha ,\alpha,\alpha ^{2},\alpha ^{4})$ leads to the 3 family SM.  
Assigning VEVs to 
$(3,1,\bar{3},1,1,1,1,1)$, $(3,1,1,\bar{3},1,1,1)$,
$(3,\bar{3},1,1,1,1)$, $(1,3,\bar{3},1,1)$
and $(1,3,\bar{3},1)$ breaks $SU(3)^8$ down to 
$SU(3)_{1235} \times SU(3)_{467} \times SU(3)_8$.  

Another option exists for 
${\bf 4}=(\alpha ,\alpha^4,\alpha^{5},\alpha^{6})$, when assigning VEVs to
$(3,\bar{3},1,1,1,1,1,1)$, $(3,\bar{3},1,1,1,1,1)$, $(3,1,1,\bar{3},1,1)$,
$(1,3,\bar{3},1,1)$ and $(1,3,1,\bar{3})$
\footnote{This SSB pathway has first been
derived by Yasmin Anstruther.}. 
These models have not been 
discussed in the literature so far and have potential interesting 
phenomenology.

\section{  ${ \cal N}=1$ and ${\cal N}=0$ chiral models for abelian product group orbifolding}

 Now let us consider abelian orbifold groups of order $o(G)\leq 12,$ that
are
not just $Z_{n}$. There are only four, but they will be sufficient to
teach
us how to deal with this type of orbifold. We will search for both
${\cal N}%
=1$ and ${\cal N}=0$ models since neither have been studied in general in
the
literature. Three groups,  
$Z_{2}\times Z_{4},$ $Z_{3}\times Z_{3}$, and $Z_{2}\times Z_{2}\times Z_{3}$ 
fit our requirements. We
have dispensed with $Z_{2}\times Z_{2}\times Z_{2}$ since all its
representations are real and it cannot produce chiral models.

First for $Z_{2}\times Z_{4},$ we can write elements as ($\alpha
^{i},\beta
^{i^{\prime }})$ where $\alpha ^{2}=1,$ and $\beta ^{4}=1.$ The
supersymmetry after orbifolding is determined by the embeddings. These
are
of the form:
\[
{\bf 4}=((\alpha ^{i},\beta ^{i^{\prime }}),(\alpha ^{j},\beta
^{j^{\prime
}}),(\alpha ^{k},\beta ^{k^{\prime }}),(\alpha ^{l},\beta ^{l^{\prime
}})).
\]
If all four entries are nontrivial ${\cal N}=0$ $SUSY$ results, if one
is
trivial, then we have ${\cal N}=1.$ We can think of the $SUSY$ breaking
as a
two step process, where we first embed the $\alpha $'$s$ in the ${\bf
4}$
and then the $\beta $'$s.$ Let us proceed this way and include only the
partition, and possibly double partition models. (As we noted above, the
nonpartition models have potential pathologies.) Thus for the $\alpha
$'$s$
we must have either ${\bf 4}_{{\bf \alpha }_{{\bf 1}}}=(-1,-1,-1,-1)$ or
$%
{\bf 4}_{{\bf \alpha }_{2}}=(1,1,-1,-1).$ The ${\bf 4}_{{\bf \alpha
}_{{\bf 1%
}}}$ results in ${\cal N}=0$ $SUSY,$ while ${\bf 4}_{{\bf \alpha }_2}$
gives
${\cal N}=2.$ We do not include trivial $Z_n$ factors {\bf 4}=(1,1,1,1) in the 
discussion, since these models contain very little new information.
[Note, for any product groups $Z_n \times Z_m$, the $\alpha$'s of $Z_n$ 
must be self conjugate in the $\bf 6$, as are the $\beta$'s of $Z_m$.
Hence, the full $\bf 6$ is self conjugate since the subgroups $Z_n $ and $ Z_m$
are orthogonal. This generalizes to more complicated products $Z_n \times Z_m \times Z_p \times ... $.]

Now for the $\beta $'$s.$ These are to be combined with the $\alpha
$'$s,$
so we must consider the ${\bf 4}_{{\bf \alpha }_{{\bf 1}}}$ and ${\bf
4}_{%
{\bf \alpha }_{{\bf 2}}}$ separately. For ${\bf 4}_{{\bf \alpha }_{{\bf
1}}},
$ the inequivalent ${\bf 4}_{{\bf \beta }}$ 's are ${\bf 4}_{{\bf \beta
}_{%
{\bf 1}}}=(\beta ,\beta ,\beta ,\beta )$ and ${\bf 4}_{\beta
_{2}}=(1,\beta
,\beta ,\beta ^{2}).$ [Models with ${\bf 4}=(1,1,\beta ^{2},\beta ^{2})$ are
uninteresting since they all are nonchiral.]
Both cases have ${\cal N}=0$ $SUSY$ since we were
already at ${\cal N}=0$ after the ${\bf 4}_{{\bf \alpha }_{{\bf 1}}}$
embedding. For ${\bf 4}_{{\bf \alpha }_{{\bf 2}}}$ we find five possible
inequivalent embeddings, again we can have ${\bf 4}_{\beta _{{\bf 1}%
}}=(\beta ,\beta ,\beta ,\beta )$ or ${\bf 4}_{\beta _{2}}=(1,\beta
,\beta
,\beta ^{2}),$ but now we can also have ${\bf 4}_{\beta _{3}}=(1,\beta
^{2},\beta ,\beta )$, ${\bf 4}_{\beta _{4}}=(\beta ,\beta ,1,\beta
^{2})$ and ${\bf 4}_{\beta _{5}}=(\beta^2 ,\beta ,1,\beta)$.
The embeddings ${\bf 4}_{\beta _{{\bf 1}}}$, ${\bf 4}_{\beta _{4}}$ 
and ${\bf 4}_{\beta _{5}}$ lead to
${\cal N%
}=0$ $SUSY$ while ${\bf 4}_{\beta _{2}}$ and  ${\bf 4}_{\beta _{3}}$
leave $%
{\cal N}=1$ $SUSY$ unbroken. A similar analysis can be carried out for
$Z_{3}%
\times Z_{3}$, and $Z_{2}\times Z_{2}\times Z_{3},$ with the obvious
generalization to a triple embedding for $Z_{2}\times Z_{2}\times
Z_{3}.$

For $Z_{3}\times Z_{3}$ there are five models. We can choose  ${\bf
4}_{{\bf %
\alpha }}=(1,\alpha ,\alpha ,\alpha )$ as the embedding of the first
$Z_{3}.$
Then the embedding of the second $Z_{3}$ can be ${\bf 4}_{{\bf \beta
}_{{\bf %
1}}}=(1,\beta ,\beta ,\beta ),$ ${\bf 4}_{{\bf \beta }_{{\bf 2}}}=(\beta
,1,\beta ,\beta ),$ ${\bf 4}_{{\bf \beta }_{{\bf 3}}}=(1,1,\beta ,\beta
^{2}),$ ${\bf 4}_{{\bf \beta }_{{\bf 4}}}=(\beta ,1,1,\beta ^{2}),$ or
${\bf %
4}_{{\bf \beta }_{{\bf 5}}}=(\beta ^{2},1,1,\beta ).$ The first and
third
result in ${\cal N}=1$ $SUSY$ models while the other three are ${\cal
N}=0.$

For $Z_{2}\times Z_{2}\times Z_{3}$ we find 9 chiral models. Rather than
belabor the details, we summarize all our results for $Z_{2}\times Z_{4},$
$Z_{3}%
\times Z_{3}$, and $Z_{2}\times Z_{2}\times Z_{3}$ in Table 5.

\section{Conclusions}

We have now completed our task of summarizing all ${\cal N}=0$ and
${\cal N}=1$  $SUSY$ chiral models of phenomenological interest
derivable from orbifolding $AdS_{5}\times S^{5}$ with abelian
orbifold group $\Gamma $ of order
$o(\Gamma )\leq 12.$ The models fall into three classes:
partition models, double partition models, and non-partition models
as determined by how the equation $i+j+k+l=sn$ is satisfied by the
embedding where $s=1$ for
partition models, $s=2$ for double partition models and $s$ is
non--integer for non-partition models.
For $Z_{n}$ orbifolds with ${\cal N}=1$ $SUSY$, there are 53
partition models, and 7 non-partition models,
and for ${\cal N}=0$ $SUSY$, we find 54 partition, 11 double
partition, and 13 non-partition
models. The non-partition models have potential pathologies if they are
to be interpreted as coming from string theory, but they still may be of
phenomenological and technical interest, so they have been included in
our classification of $Z_{n}$ models. See also the related discussions in \cite{Pickering:2001aq} and
\cite{Kakushadze:2000mc}.

The non--$Z_{n}$  abelian product groups of interest (we only consider
partition models here) with $o(\Gamma )\leq 12$ are
$Z_{2}\times Z_{4}$ with five  ${\cal N}=0$ and two
${\cal N}=1$ chiral models; $Z_{3}\times Z_{3}$ with three ${\cal N}=0$ and two ${\cal N}=1$ chiral
models, and $Z_{2}\times Z_{2}\times Z_{3}$ with seven ${\cal N}=0$ and two
${\cal N}=1$ chiral models.

We have explored the relation to the SM and MSSM in some detail only for
$Z_{n}$
models with $o(\Gamma )\leq 7$, but have only given a few examples with
$o(\Gamma)>7,$ and have
indicated how to build abelian orbifold models for any $o(\Gamma )$. 
Two $Z_8$ models have been introduced, which can lead to the right
Weinberg angle, when broken down to the SM.
We
hope our results will be
useful to model builders and phenomenologists alike.

\emph{Acknowledgments.---}  
We thank Yasmin Anstruther for working out several $Z_8$ SSB pathways.
The work of TK was supported by U.S. 
DoE grant number
DE-FG05-85ER40226. HP
was supported by the Bundesministerium f\"ur Bildung und Forschung 
(BMBF, Bonn, Germany) under the contract number 05HT1WWA2.

\newpage

\begin{tabular}{|c||c|c|c|}
\hline
$n$ & {\bf 4} & $\chi /N^{2}$ & comment \\ \hline\hline
3 & $({\bf 1},\alpha ,\alpha ,\alpha )$ & 9 & $i+j+k=3;$ one model $%
(i=j=k=1) $ \\ \hline
3 & $({\bf 1},\alpha ,\alpha ,\alpha ^{2})^{*}$ & 3 &  \\ \hline
4 & $({\bf 1},\alpha ,\alpha ,\alpha ^{2})$ & 8 & $i+j+k=4;$ one model \\ 
\hline
5 & $({\bf 1},\alpha ^{i},\alpha ^{j},\alpha ^{k})$ & 15 & $i+j+k=5;$ 2
models \\ \hline
6 & $({\bf 1},\alpha ^{i},\alpha ^{j},\alpha ^{k})$ & 12 & $i+j+k=6;$ 3
models \\ \hline
6 & $({\bf 1},\alpha ,\alpha ^{2},\alpha ^{4})^{*}$ & 6 &  \\ \hline
6 & $({\bf 1},\alpha ^{2},\alpha ^{2},\alpha ^{4})^{*}$ & 6 &  \\ \hline
7 & $({\bf 1},\alpha ^{i},\alpha ^{j},\alpha ^{k})$ & 21 & $i+j+k=7;$ 4
models \\ \hline
8 & $({\bf 1},\alpha ^{i},\alpha ^{j},\alpha ^{k})$ & $\leq 24$ & $i+j+k=8;$
5 models \\ \hline
9 & $({\bf 1},\alpha ^{i},\alpha ^{j},\alpha ^{k})$ & 27 & $i+j+k=9;$ 7
models \\ \hline
9 & $({\bf 1},\alpha ,\alpha ^{4},\alpha ^{7})^{*}$ & 27 &  \\ \hline
9 & $({\bf 1},\alpha ^{3},\alpha ^{3},\alpha ^{6})^{*}$ & 9 &  \\ \hline
10 & $({\bf 1},\alpha ^{i},\alpha ^{j},\alpha ^{k})$ & 30 & $i+j+k=10;$ 8
models \\ \hline
11 & $({\bf 1},\alpha ^{i},\alpha ^{j},\alpha ^{k})$ & 33 & $i+j+k=11;$ 10
models \\ \hline
12 & $({\bf 1},\alpha ^{i},\alpha ^{j},\alpha ^{k})$ & $\leq 36$ & $%
i+j+k=12; $ 12 models \\ \hline
12 & $({\bf 1},\alpha ^{2},\alpha ^{4},\alpha ^{8})^{*}$ & 12 &  \\ \hline
12 & $({\bf 1},\alpha ^{4},\alpha ^{4},\alpha ^{8})^{*}$ & 12 &  \\ \hline
\end{tabular}
\\
\\
Table 1: 
All ${\cal N}=1$ chiral $Z_{n}$ orbifold models with $n\leq 12.$ Three of the $%
n=8$ models have $\chi /N^{2}=24;$ the other two have $\chi /N^{2}=16.$ Of
the 12 models with $i+j+k=12,$ three have models $\chi /N^{2}=24$ and the
other nine have $\chi /N^{2}=36.$ Of the 60 models 53 are partition models,
while the remaining 7 models that do not satisfy $i+j+k=n$, are marked with
an asterisk (*).

\newpage


\begin{tabular}{|c||c|c|c|c|}
\hline
$M_{1111}^{4}(F)$ & 1 & $\alpha $ & $\alpha ^{2}$ & $\alpha ^{3}$ \\ 
\hline\hline
$1$ &  & $\times ^{4}$ &  &  \\ \hline
$\alpha $ &  &  & $\times ^{4}$ &  \\ \hline
$\alpha ^{2}$ &  &  &  & $\times ^{4}$ \\ \hline
$\alpha ^{3}$ & $\times ^{4}$ &  &  &  \\ \hline
\end{tabular}
\\
\\
Table 2: Fermion content for the model $M_{1111}^{4}.$ The $\times ^{4}$
entry at the (1,$\alpha)$ position means the model contains $4(3,\bar{3}%
,1,1)$ of $SU^{4}(3),$ etc. Hence, the fermions in this table are 4[$(3,\bar{%
3},1,1)+(1,3,\bar{3},1)+(1,1,3,\bar{3})+(\bar{3},1,1,3)$]. Diagonal entries
do not occur in this model but, if they did, an $\times $ at say ($\alpha ^{2}$,$\alpha
^{2})$ would correspond to $(1,8+1,1,1)$, etc. See models below.

\bigskip

\bigskip

\begin{tabular}{|c||c|c|c|c|}
\hline
$M_{1111}^{4}(S)$ & 1 & $\alpha $ & $\alpha ^{2}$ & $\alpha ^{3}$ \\ 
\hline\hline
$1$ &  &  & $\times ^{6}$ &  \\ \hline
$\alpha $ &  &  &  & $\times ^{6}$ \\ \hline
$\alpha ^{2}$ & $\times ^{6}$ &  &  &  \\ \hline
$\alpha ^{3}$ &  & $\times ^{6}$ &  &  \\ \hline
\end{tabular}
\\
\\
Table 3: Scalar content of the model $M_{1111}^{4}.$

\bigskip


\newpage

\begin{tabular}{|c||c|c|c|}
\hline
$n$ & {\bf 4} & $\chi /N^{2}$ & comment \\ \hline\hline
4 & $(\alpha ,\alpha ,\alpha ,\alpha )$ & 16 & $i+j+k+l=3;$ one model $%
(i=j=k=l=1)$ \\ \hline
4 & $(\alpha ,\alpha ,\alpha ,\alpha ^{3})^*$ & 8 & nonpartition model \\ 
\hline
5 & $(\alpha ^{i},\alpha ^{j},\alpha ^{k},\alpha ^{l})$ & 20 & $i+j+k+l=5;$
1 models \\ \hline
6 & $(\alpha ^{i},\alpha ^{j},\alpha ^{k},\alpha ^{l})$ & $\leq $24 & $%
i+j+k+l=6;$ 2 models \\ \hline
6 & $(\alpha ,\alpha ,\alpha ^{3},\alpha ^{5})^{*}$ & 6 
& nonpartition \\ \hline
6 & $(\alpha ,\alpha ^{2},\alpha ^{3},\alpha ^{5})^{*}$ & 6 & 
nonpartition \\ \hline
6 & $(\alpha ,\alpha ^{3},\alpha ^{4},\alpha ^{4})$ & 24 & double partition
\\ \hline
7 & $(\alpha ^{i},\alpha ^{j},\alpha ^{k},\alpha ^{l})$ & 28 & $i+j+k+l=7;$
3 models \\ \hline
8 & $(\alpha ^{i},\alpha ^{j},\alpha ^{k},\alpha ^{l})$ & $\leq $ $32$ & $%
i+j+k+l=8;$ 5 models \\ \hline
8 & $(\alpha ,\alpha ^{2},\alpha ^{3},\alpha ^{6})^{*}$ & 16 & nonpartition
\\ \hline
8 & $(\alpha ^{2},\alpha ^{2},\alpha ^{2},\alpha ^{6})^{*}$ & 16 & analog of 
$Z_{4}$ $(\alpha ,\alpha ,\alpha ,\alpha ^{3})$ model \\ \hline
8 & $(\alpha ,\alpha ^{4},\alpha ^{5},\alpha ^{6})$ & 32 & double partition
\\ \hline
9 & $(\alpha ^{i},\alpha ^{j},\alpha ^{k},\alpha ^{l})$ & 36 & $i+j+k+l=9;$
7 models \\ \hline
9 & $(\alpha ,\alpha ^{3},\alpha ^{4},\alpha ^{7})^{*}$ & 36 & nonpartition
\\ \hline
9 & $(\alpha ,\alpha ^{4},\alpha ^{6},\alpha ^{7})$ & 36 & double partition
\\ \hline
10 & $(\alpha ^{i},\alpha ^{j},\alpha ^{k},\alpha ^{l})$ & $\leq 40$ & $%
i+j+k+l=10;$ 9 models \\ \hline
10 & $(\alpha ,\alpha ^{3},\alpha ^{8},\alpha ^{8})$ & 40 & double partition
\\ \hline
10 & $(\alpha ,\alpha ^{5},\alpha ^{6},\alpha ^{8})$ & 40 & double partition
\\ \hline
11 & $(\alpha ^{i},\alpha ^{j},\alpha ^{k},\alpha ^{l})$ & 44 & $i+j+k+l=11;$
11 models \\ \hline
12 & $(\alpha ^{i},\alpha ^{j},\alpha ^{k},\alpha ^{l})$ & $\leq $ $48$ & $%
i+j+k+l=12;$ 15 models \\ \hline
12 & $(\alpha ,\alpha ^{4},\alpha ^{9},\alpha ^{10})$ & 48 & double partition
\\ \hline
12 & $(\alpha ,\alpha ^{5},\alpha ^{9},\alpha ^{9})$ & 48 & double partition
\\ \hline
12 & $(\alpha ,\alpha ^{6},\alpha ^{7},\alpha ^{10})$ & 48 & double partition
\\ \hline
12 & $(\alpha ,\alpha ^{6},\alpha ^{8},\alpha ^{9})$ & $36$ & double
partition \\ \hline
12 & $(\alpha ,\alpha ^{7},\alpha ^{8},\alpha ^{8})$ & 48 & double partition
\\ \hline
12 & $(\alpha ^{2},\alpha ^{6},\alpha ^{8},\alpha ^{8})$ & 36 & double
partition \\ \hline
12 & $(\alpha ,\alpha ,\alpha ^{5},\alpha ^{9})^*$ & 48 & nonpartition \\ 
\hline
12 & $(\alpha ,\alpha ^{3},\alpha ^{5},\alpha ^{9})^{*}$ & 24 & nonpartition
\\ \hline
12 & $(\alpha ,\alpha ^{3},\alpha ^{7},\alpha ^{11})^{*}\ $ & 24 & 
nonpartition \\ \hline
12 & $(\alpha ,\alpha ^{5},\alpha ^{5},\alpha ^{9})^{*}$ & 48 & nonpartition
\\ \hline
12 & $(\alpha ^{2},\alpha ^{2},\alpha ^{6},\alpha ^{10})^{*}$ & 12 & 
nonpartition \\ \hline
12 & $(\alpha ^{2},\alpha ^{3},\alpha ^{4},\alpha ^{9})^{*}$ & 24 & 
nonpartition \\ \hline
12 & $(\alpha ^{2},\alpha ^{4},\alpha ^{6},\alpha ^{10})^{*}$ & 24 & 
nonpartition \\ \hline
12 & $(\alpha ^{3},\alpha ^{3},\alpha ^{3},\alpha ^{9})^{*}$ & 24 & 
nonpartition \\ \hline
\end{tabular}
\\
\\
Table 4. 
All chiral ${\cal N}=0,$ $Z_{n}$ orbifold models with $n\leq12.$  
The 13 non--partition models are marked with an asterisk(*).
For further explanations see text.
\newpage

\bigskip
\begin{tabular}{|c||c|c|c|}
\hline
$Group$ & {\bf 4} & $\chi /N^{2}$ & ${\cal N}$ \\ \hline\hline
$Z_{2}\times Z_{4}$ & $(-1,-1,-1,-1)\times (\beta ,\beta ,\beta ,\beta
)$ &
32 & $0$ \\ \hline
$Z_{2}\times Z_{4}$ & $(-1,-1,-1,-1)\times ({\bf 1},\beta ,\beta ,\beta
^{2})
$ & 16 & 0 \\ \hline
$Z_{2}\times Z_{4}$ & $(1,1,-1,-1)\times (\beta ,\beta ,\beta ,\beta )$
& 32
& $0$ \\ \hline
$Z_{2}\times Z_{4}$ & $(1,1,-1,-1)\times ({\bf 1},\beta ,\beta ,\beta
^{2})$
& 16 & $1$ \\ \hline
$Z_{2}\times Z_{4}$ & $(1,1,-1,-1)\times ({\bf 1},\beta ^{2},\beta
,\beta )$
& 16 & $1$ \\ \hline
$Z_{2}\times Z_{4}$ & $(1,1,-1,-1)\times (\beta ,\beta ,1,\beta ^{2})$ &
16
& $0$ \\ \hline
$Z_{2}\times Z_{4}$ & $(1,1,-1,-1)\times (\beta ,\beta ^{2} ,1,\beta)$ &
16
& $0$ \\ \hline
$Z_{3}\times Z_{3}$ & $({\bf 1},\alpha ,\alpha ,\alpha )\times (1,\beta
,\beta ,\beta )$ & 27 & $1$ \\ \hline
$Z_{3}\times Z_{3}$ & $({\bf 1},\alpha ,\alpha ,\alpha )\times (\beta
,1,\beta ,\beta )$ & 36 & $0$ \\ \hline
$Z_{3}\times Z_{3}$ & $({\bf 1},\alpha ,\alpha ,\alpha )\times
(1,1,\beta
,\beta ^{2})$ & 18 & 1 \\ \hline
$Z_{3}\times Z_{3}$ & $({\bf 1},\alpha ,\alpha ,\alpha )\times (\beta
,1,1,\beta ^{2})$ & 36 & 0 \\ \hline
$Z_{3}\times Z_{3}$ & $({\bf 1},\alpha ,\alpha ,\alpha )\times (\beta
^{2},1,1,\beta )$ & 36 & $0$ \\ \hline
%
$Z_{2}\times Z_{2}\times Z_{3}$ & $(1,1,-1,-1)\times (1,1,-1,-1)\times %

(1,\gamma ,\gamma ,\gamma )$ & $48$ & $1$ \\ \hline
$Z_{2}\times Z_{2}\times Z_{3}$ & $(1,1,-1,-1)\times (-1,1,1,-1)\times %

(1,\gamma ,\gamma ,\gamma )$ & $48$ & $0$ \\ \hline
$Z_{2}\times Z_{2}\times Z_{3}$ & $(1,1,-1,-1)\times (-1,-1,-1,-1)\times
(1,\gamma ,\gamma ,\gamma )$ & $48$ & 0 \\ \hline
$Z_{2}\times Z_{2}\times Z_{3}$ & $(-1,-1,1,1)\times (-1,-1,1,1)\times %

(1,\gamma ,\gamma ,\gamma )$ & $48$ & 0 \\ \hline
$Z_{2}\times Z_{2}\times Z_{3}$ & $(-1,-1,1,1)\times (-1,-1,-1,-1)\times
(1,\gamma ,\gamma ,\gamma )$ & $48$ & 0 \\ \hline
$Z_{2}\times Z_{2}\times Z_{3}$ & $(1,1,-1,-1)\times
(-1,-1,1,1)\times %
(1,\gamma ,\gamma ,\gamma )$ & $48$ & 0 \\ \hline
$Z_{2}\times Z_{2}\times Z_{3}$ & $(1,1,-1,-1)\times
(1,-1,-1,1)\times %
(1,\gamma ,\gamma ,\gamma )$ & $48$ & 1 \\ \hline
$Z_{2}\times Z_{2}\times Z_{3}$ & $(-1,1,1,-1)\times
(-1,1,-1,1)\times %
(1,\gamma ,\gamma ,\gamma )$ & $48$ & 0 \\ \hline
$Z_{2}\times Z_{2}\times Z_{3}$ & $(-1,-1,-1,-1)\times
(-1,-1,-1,-1)\times %
(1,\gamma ,\gamma ,\gamma )$ & $48$ & 0 \\ \hline
\end{tabular}
\\
\\
Table 5.: All chiral ${\cal N}=0$ and ${\cal N}=1$ $SUSY$ partition models for product
orbifolding groups $Z_{2}\times Z_{4},$ $Z_{3}\times Z_{3}$, and $Z_{2}%
\times Z_{2}\times Z_{3}$, where the embedding is nontrivial in all factors. Our notation is:
${\bf 4}=
((\alpha ^{i}),(\alpha ^{j}),(\alpha ^{k}),(\alpha ^{l}))\times
((\beta ^{i^{\prime }}),(\beta^{j^{\prime
}}),(\beta ^{k^{\prime }}),(\beta ^{l^{\prime
}}))
=((\alpha ^{i},\beta ^{i^{\prime }}),(\alpha ^{j},\beta
^{j^{\prime
}}),(\alpha ^{k},\beta ^{k^{\prime }}),(\alpha ^{l},\beta ^{l^{\prime
}}))$, etc.

\end{document}